\documentclass[12pt]{iopart} 
 
\usepackage{graphicx,epsfig} 
\newcommand{ \be }{\begin{equation}} 
\newcommand{ \ee }{\end{equation}} 
\newcommand{ \bea }{\begin{eqnarray}} 
\newcommand{ \eea }{\end{eqnarray}}

\usepackage{color}

\begin{document} 
 
\title[Centrality dependence of hyperon global polarization in Au+Au collisions]{Centrality dependence of hyperon global polarization in Au+Au collisions at RHIC} 
 
\author{I Selyuzhenkov for the STAR Collaboration} 
 
\address{Wayne State University, 666 W Hancock, Detroit MI 48201, USA} 
\ead{Ilya.Selyuzhenkov@wayne.edu} 
 
\begin{abstract}
We present the centrality dependence of $\Lambda$ and $\bar\Lambda$ hyperon global polarization in Au+Au collisions
at $\sqrt{s_{NN}}=62$~GeV and $200$~GeV measured with the STAR detector at RHIC.
Within the precision of the measurement, we observe no centrality dependence of $\Lambda$ and $\bar\Lambda$
hyperon global polarization and within our acceptance it is consistent with zero.
Different sources of systematic uncertainties (feed down effects, spin precession) are discussed and estimated.
The obtained upper limit,  $|P_{\Lambda,\bar\Lambda}| \leq 0.02$, is
compared to theoretical predictions discussed recently in literatures.
\end{abstract}
 
\section{Introduction}
\label{Introduction} The system created in non-central relativistic nucleus-nucleus
collisions possesses large orbital angular momentum. One of the most interesting and
important phenomena predicted to occur in such a system is global polarization of the
system~\cite{LiangPRL94, Voloshin0410089, Liang0411101}. This phenomenon manifests
itself in the polarization of produced secondary particles along the direction of the 
system's angular momentum. Measurement of global polarization may provide valuable insight into the
evolution of the system, the mechanism for hadronization, and the origin of hadronic spin 
preferences.  The orbital angular momentum of the system may be transformed into
a preference in global spin orientation of particles
by spin-orbit coupling at various stages of the system
evolution. This can happen at the partonic level, while the system evolves as an ensemble of 
deconfined polarized quarks. Polarization of hadrons produced secondarily could also
be acquired via hadron re-scattering at a later hadronic stage. An example of 
transformation from system orbital momentum to the global polarization of produced $\rho$-mesons, which is due to 
the pion re-scattering, is discussed in~\cite{Voloshin0410089}. 
One specific scenario for the spin-orbit transformation via the polarized quark phase is 
discussed in~\cite{LiangPRL94}. There, it is argued that parton interactions in 
non-central relativistic nucleus-nucleus collisions lead first to the global polarization 
of the produced quarks. The magnitude for this global quark polarization at RHIC 
(Relativistic Heavy Ion Collider) energies was estimated to be as high as thirty percent.
Assuming that the strange and non 
strange quark polarizations ($P_s$ and $P_q$, respectively) are equal, in the particular case of the 
`exclusive' parton recombination scenario~\cite{LiangPRL94}, the global 
polarization $P_H$ for $\Lambda$, $\Sigma$, and $\Xi$ hyperons appears to be similar to 
that for quarks: $P_H = P_q \simeq 0.3$. Recently more realistic
calculations~\cite{Liang:Xian2006} of the global quark polarization were performed within a model based 
on the HTL (Hard Thermal Loop) gluon propagator. The resulting hyperon polarization was 
predicted to be in the range from $-0.03$ to $0.15$, depending on the temperature of the
QGP formed.
 
In this paper we present the centrality dependence of $\Lambda$ and $\bar\Lambda$ hyperon global 
polarization in Au+Au collisions   at $\sqrt{s_{NN}}$=62 and 
200~GeV measured with the STAR (Solenoidal Tracker At RHIC) detector. 
 
\section{Global polarization of hyperons} 
In the hyperon global polarization measurement we use the observable derived in~\cite{Selyuzhenkov:2005xa, Selyuzhenkov:2006fc}: 
\begin{eqnarray} 
\label{GlobalPolarizationObservable} 
P_{H}~=~\frac{8}{\pi\alpha_H}\langle \sin \left( \phi^*_p 
- \Psi_{RP}\right)\rangle~~. 
\end{eqnarray} 
where $P_{H}$ is the hyperon global polarization, 
 $\Psi_{RP}$  is the reaction plane angle, and $\alpha_H$ is the hyperon decay parameter. 
$\phi^*_p$ is the azimuthal angle of the 3-momentum of hyperon's decay product, measured in hyperon's rest frame.
Angle brackets in this equation denote an average over the solid angle of the hyperon's 
decay product 3-momentum in the hyperon's rest frame and over all directions of the system 
orbital momentum {\boldmath $L$}, or, in other words, over all possible orientations of 
the reaction plane. 
Note that in Eq.~\ref{GlobalPolarizationObservable}, perfect detector acceptance is assumed. 
A detailed study~\cite{Selyuzhenkov:2006tj} of detector acceptance effects
shows that related systematic uncertainty is less than $20\%$.
 
The hyperon reconstruction procedure used in this analysis is similar to that
in~\cite{Adler:2002pb,Cai:2005ph,Takahashi:2005pq}. $\Lambda$ and $\bar\Lambda$ particles 
were reconstructed from their weak decay topology, $\Lambda \to p \pi^- $ and $\bar\Lambda 
\to \bar p \pi^+ $, using charged tracks measured in the STAR main TPC 
(Time Projection Chamber)~\cite{Anderson:2003ur}. The corresponding decay 
parameter is $\alpha_{\Lambda}^{-} = - \alpha_{\bar\Lambda}^{+} = 0.642\pm0.013$~\cite{Eidelman:2004wy}.
In this analysis hyperon candidates with invariant 
mass within the window $1.11<m_{\Lambda, \bar\Lambda}<1.12$ $\rm GeV/c^2$ are used. 
The statistics for $\bar\Lambda$--hyperons is smaller than that for $\Lambda$--hyperons 
by 40\% (20\%) for Au+Au collisions at $\sqrt{s_{NN}}$=62~GeV (200~GeV). 
The background contribution is estimated by fitting the 
invariant mass distribution with the sum of a Gaussian and a 3rd-order polynomial function, 
and is found to be less than 8\%. 
 
Collision centrality was 
defined using the total charged particle multiplicity within a pseudorapidity window of 
$|\eta|<0.5$. The charged particle multiplicity distribution was divided into nine 
centrality bins (classes): {\mbox{0-5\%} (most central collisions)}, \mbox{5-10\%}, \mbox{10-20\%}, 
\mbox{20-30\%}, \mbox{30-40\%}, \mbox{40-50\%}, \mbox{50-60\%}, \mbox{60-70\%}, and \mbox{70-80\%} of the total hadronic 
inelastic cross section for Au+Au collisions. 
 
The reaction plane angle in Eq.~\ref{GlobalPolarizationObservable} is estimated by 
calculating the so-called event plane flow vector $Q_{EP}$. 
This implies the necessity to correct the final results by the reaction plane resolution 
$R_{EP}$~\cite{Voloshin:1994mz,Barrette:1996rs,Poskanzer:1998yz}. 
The global polarization measurement requires the knowledge of the direction of the system 
orbital momentum {\boldmath $L$}, hence, of the first-order event plane vector. 
In this paper, the first-order event plane vector was determined from two STAR Forward TPCs~\cite{Ackermann:2002yx}, 
which span a pseudorapidity region $2.7 < |\eta| < 3.9$.
Charged particle tracks with transverse momentum 
$0.15 < p_t < 2.0$ GeV are used to define the event plane vector. 
It was found in this analysis that the event plane vector defined with the particles 
measured in the Forward 
TPCs is reliable within the centrality region \mbox{0-80\%} for Au+Au 
collisions at $\sqrt{s_{NN}}$=62~GeV. 
With higher multiplicity at  $\sqrt{s_{NN}}$=200~GeV, 
saturation effects in the Forward TPCs 
for the most central collisions become evident, 
and the estimated reaction plane angle is unreliable. 
Due to this effect, the centrality region used for the 
$\Lambda$ ($\bar\Lambda$) hyperon global polarization measurement  in Au+Au collisions 
at $\sqrt{s_{NN}}$=200~GeV is limited to \mbox{20-70\%}. 
 
The measured hyperons consist of primordial 
$\Lambda$ ($\bar\Lambda$) and feed down from multiply strange hyperons ($\Xi^0$ and 
$\Omega$) and $\Sigma^0$ decays, and also from short-lived resonances decaying via strong 
interactions. Under the assumption that the global polarization has the same value for 
$\Lambda$ and $\Sigma^0$~\cite{LiangPRL94}, we estimate the relative contribution from 
$\Sigma^0$ to the extracted global polarization of the $\Lambda$ hyperons to be $\leq 
30$\%.  This estimate takes into account an average polarization transfer from $\Sigma^0$ 
to $\Lambda$ of $-1/3$~\cite{PhysRev.140.B668,Armenteros:1970eg} (this value can be 
affected by non-uniform acceptance of the daughter $\Lambda$). It furthermore allows for 
the $\Sigma^0/\Lambda$ production ratio to be 2-3 times higher for Au+Au collisions than 
the value (15\%) measured~\cite{VanBuren:2005px} for d+Au. Based on the results~\cite{Adams:2006ke},
the contribution of feed-downs from multiply strange hyperons ($\Xi$, 
$\Omega$) is estimated to be less than 15\%. The effect of feed-downs to $\Lambda$ 
($\bar\Lambda$) from strongly decaying resonances has not been measured with the STAR 
detector. Calculations from a string fragmentation model~\cite{Pei:1997xt} suggest that in $pp$ 
collisions the fraction of direct hyperons is about 27\% for $\Lambda$ and 15\% for 
$\bar\Lambda$, respectively.
The global polarization measurement could also conceivably be affected by hyperon spin 
precession in the strong magnetic field within the TPC. 
The effect of the spin precession on the global polarization measurements is found to be negligible. 
The overall relative uncertainty in the $\Lambda$ 
($\bar\Lambda$) hyperon global polarization measurement due to detector effects is 
estimated to be less than a factor of 2. 
 
Results of the measurement for the $\Lambda$ global polarization as 
a function of hyperon transverse momentum and pseudorapidity were presented in~\cite{Selyuzhenkov:2006fc}. 
\begin{figure}[h] 
\includegraphics[width=0.5\textwidth]{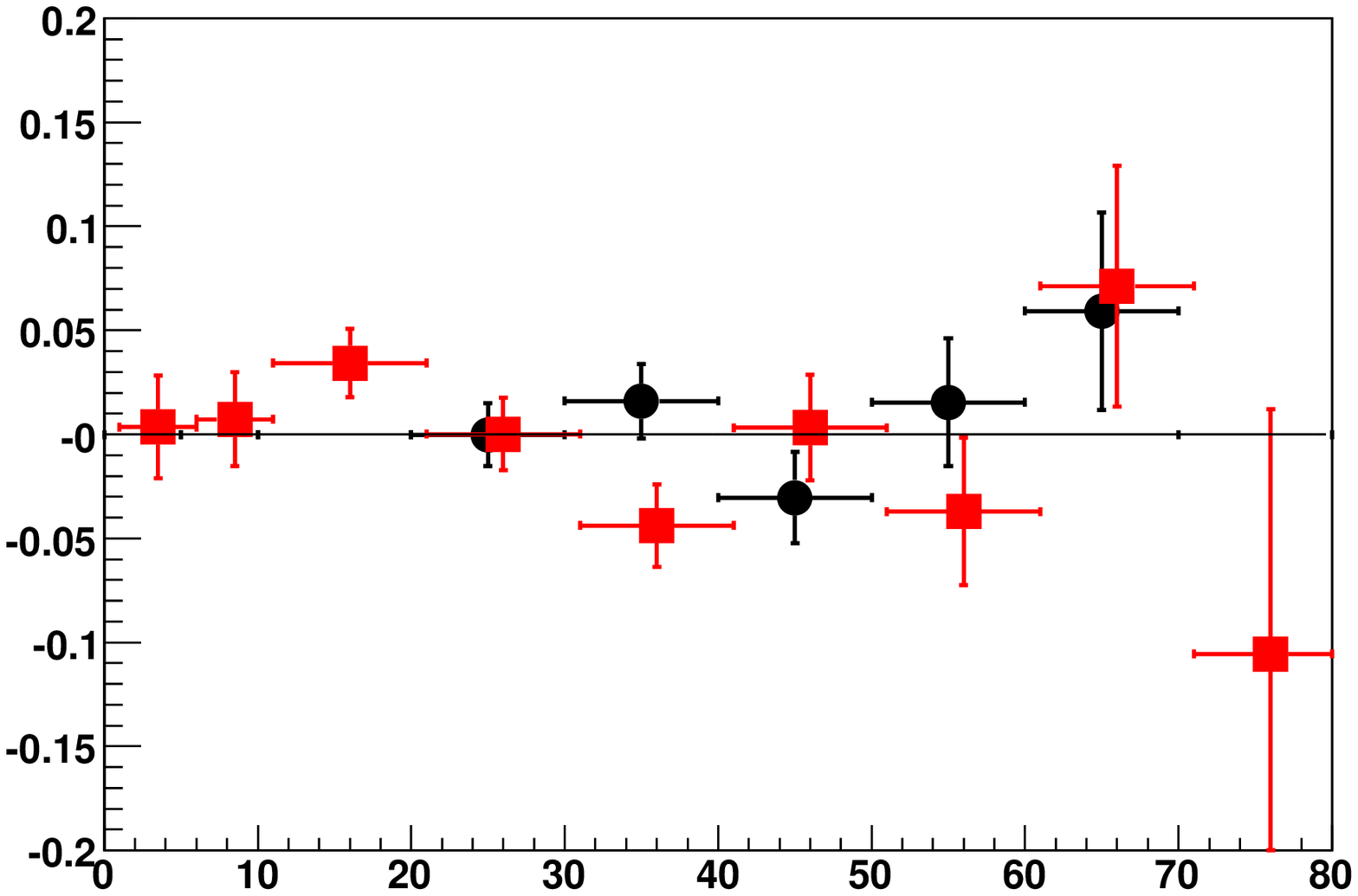}%
\includegraphics[width=0.5\textwidth]{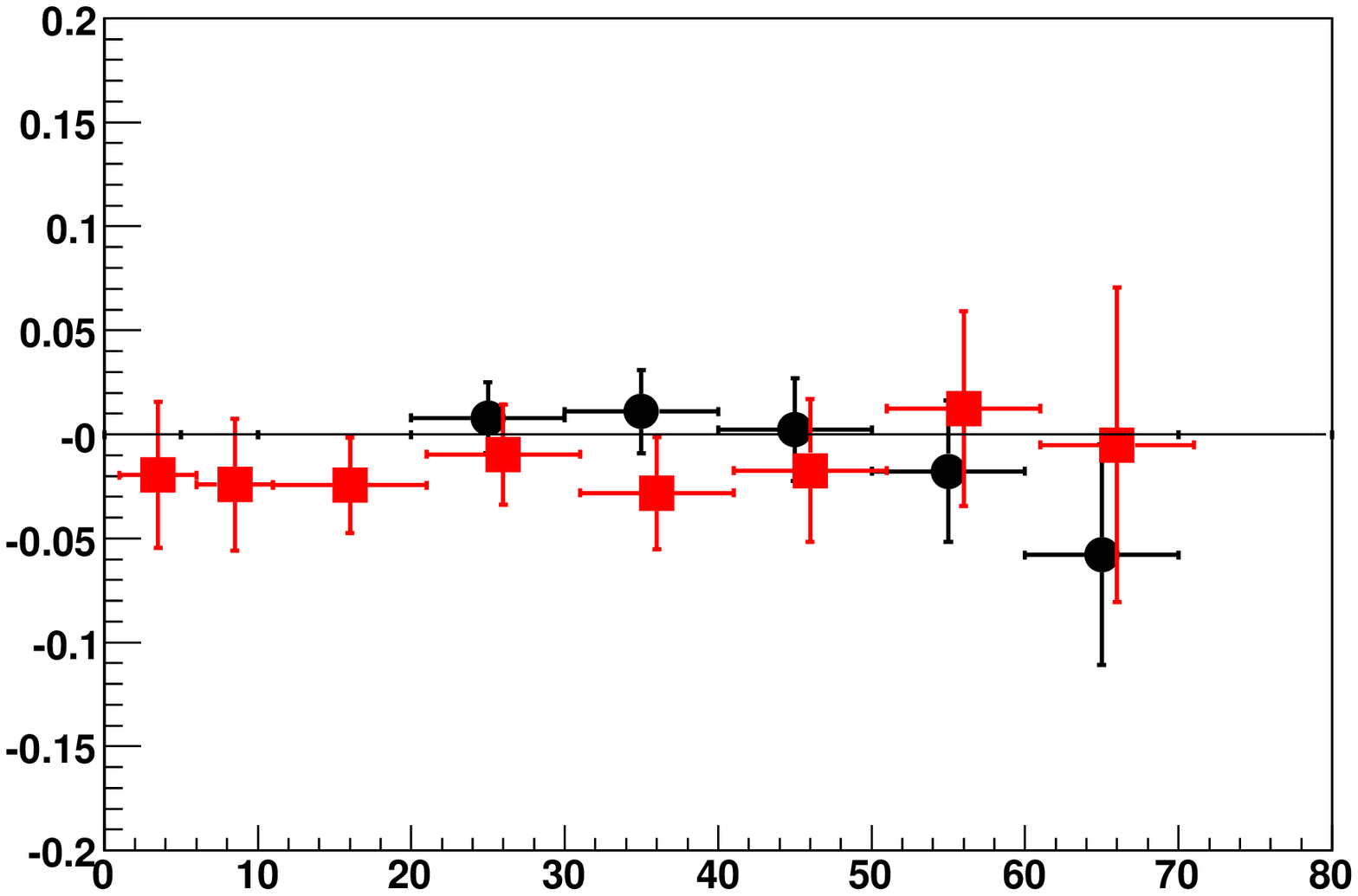} 
\put(-230,70){\rotatebox{90}{$P_{\bar\Lambda}$}} 
\put(-120,-5){$\sigma$ (\%)} 
\put(-365,120){{\bf STAR Preliminary}} 
\put(-452,70){\rotatebox{90}{$P_{\Lambda}$}} 
\put(-350,-5){$\sigma$ (\%)}%
\put(-142,120){{\bf STAR Preliminary}} 
\caption{\label{hyperonGlobalPolarization_sigma} 
(Color online) Global polarization of $\Lambda$ (left) and $\bar\Lambda$ (right) hyperons as a function of centrality 
given as fraction of the total inelastic hadronic cross section. 
Black circles show the results for Au+Au collisions 
at $\sqrt{s_{NN}}$=200~GeV (centrality region \mbox{20-70\%}) and 
red squares indicate the results for Au+Au collisions at 
$\sqrt{s_{NN}}$=62~GeV (centrality region \mbox{0-80\%}).
Only statistical errors are shown.}
\end{figure} 
Figure~\ref{hyperonGlobalPolarization_sigma} presents the $\Lambda$ and $\bar\Lambda$ hyperon global 
polarization as a function of centrality given as a fraction of the total inelastic 
hadronic cross section. Black 
circles show the result of the measurement for Au+Au collisions at 
$\sqrt{s_{NN}}$=200~GeV. Red squares indicate the result of a similar measurement for 
Au+Au collisions at $\sqrt{s_{NN}}$=62~GeV. 
Within uncertainties we observe no
centrality-dependence of the $\Lambda$  and $\bar\Lambda$ global polarization
and the obtained results are consistent with zero.
 
\section{Conclusion} 
\label{Conclusion} 
The $\Lambda$ and $\bar\Lambda$ hyperon global polarization as a function of centrality 
has been measured in Au+Au 
collisions at center of mass energies $\sqrt{s_{NN}}$=62 and 200~GeV with the STAR detector at RHIC. 
Within uncertainties we observe no 
centrality-dependence of $\Lambda$  and $\bar\Lambda$ global polarization. 
Combining results of this measurement and those from~\cite{Selyuzhenkov:2006fc}, 
an upper limit of $|P_{\Lambda,\bar\Lambda}| \leq 0.02$ for the global 
polarization of $\Lambda$ and $\bar\Lambda$ hyperons within STAR's acceptance is 
obtained. The obtained upper limit is far below the few tens of percent values discussed 
in~\cite{LiangPRL94}, but it falls within the predicted region from the more realistic 
calculations~\cite{Liang:Xian2006} based on the HTL (Hard Thermal Loop) model.

\end{document}